\documentclass[final]{svjour3}
\pdfoutput=1
\usepackage[pdftex]{graphicx}
\usepackage{bm} 
\usepackage{amsfonts}
\usepackage{amsmath}
\usepackage{graphicx}
\usepackage{rotating}
\usepackage{amssymb}
\usepackage{mathptmx}
\usepackage[numbers]{natbib}
\makeatletter
\journalname{Journal of Low Temperature Physics}

\bibpunct{[}{]}{,}{n}{}{,}

\begin{document}

\newcommand{\hdblarrow}{H\makebox[0.9ex][l]{$\downdownarrows$}-}
\titlerunning{Development of Transition-Edge Sensor X-ray Microcalorimeter Linear Array...}
\title{Development of Transition-Edge Sensor X-ray Microcalorimeter Linear Array for Compton Scattering and Energy Dispersive Diffraction Imaging}
\author{U.~Patel$^{1}$ \and R.~Divan$^{2}$ \and L.~Gades$^{1}$ \and T.~Guruswamy$^{1}$ \and D.~Yan$^{1,3}$ \and O.~Quaranta$^{1}$ \and A.~Miceli$^{1}$}
\institute{$^{1}$ Advanced Photon Source, Argonne National Laboratory, Lemont, IL 60585, USA\\ $^{2}$ Center for Nanoscale Materials, Argonne National Laboratory, Lemont, IL 60585, USA\\ $^{3}$ Applied Physics Program, Northwestern University, Evanston, IL 60208, USA\\
\email{upatel@anl.gov}}

\maketitle

\begin{abstract}

We present a strip transition-edge sensor microcalorimeter linear array detector developed for energy dispersive X-ray diffraction imaging and Compton scattering applications. The prototype detector is an array of 20 transition-edge sensors with absorbers in strip geometry arranged in a linear array. We discuss the fabrication steps needed to develop this array including Mo/Cu bilayer, Au electroplating, and proof-of-principle fabrication of long strips of SiN$_{\rm{x}}$ membranes. We demonstrate minimal unwanted effect of strip geometry on X-ray pulse response, and show linear relationship of 1/pulse height and pulse decay times with absorber length. For the absorber lengths studied, our preliminary measurements show energy resolutions of 40 eV to 180 eV near 17 keV. Furthermore, we show that the heat flow to the cold bath is nearly independent of the absorber area and depends on the SiN$_{\rm{x}}$ membrane geometry.


\end{abstract}

\section{Introduction}

Instruments based on arrays of transition-edge sensors (TESs) have been deployed for X-ray and gamma-ray spectroscopy with demonstrated energy resolution of $\sim\,$~50~eV at $\sim\,$~100 keV, and a collection area comparable to traditional High-Purity Germanium (HPGe) detectors \cite{Ullom15}. Absorbers used for these applications typically use thick Sn absorbers placed on TESs with a micromanipulator. On the other hand, microfabrication techniques in conjuction with development of electroplating of thick absorbers allow scalability to large arrays with good control over absorber geometry. Recently, we presented a design optimization for a strip TES X-ray microcalorimeter array suitable for energy dispersive X-ray diffraction (EDXRD) imaging and Compton scattering applications. These high-energy experiments are typically performed with a hard X-ray energy of $\sim\,$~100~keV, which demand a high X-ray stopping power in the detector. Our simulations showed that a strip TES detector can improve the energy resolution by an order of magnitude compared to the traditional Ge detector used in EDXRD \citep*{Yan19}. This level of improvement is needed to reach a comparable d-spacing resolution to angle dispersive diffraction $\sim\,0.04$~\AA~for EDXRD, and a momentum resolution of $\sim\,$~0.1~a.u for Compton scattering. Both of these applications benefit from an array composed of parallel strips similar to segmented Ge detectors used in EDXRD~\cite{Rumaiz18}. A strip TES with linear array would allow spatial resolution enabling efficient energy dispersive tomographic imaging experiments \cite{Stock17}. In this report, we describe the first steps taken to realize actual devices at the Advanced Photon Source (APS). Furthermore, we report on design, fabrication, and characterization of the effect of strip geometry on X-ray pulse response and report energy resolution measurements of strip TES detectors with $\sim\,$$10~\mu\text{m}$ thick Au absorbers.

\section{TES Strip Array Design and Development}

\subsection{TES Strip Array Design}

\begin{figure}[htbp]
\begin{center}
\includegraphics[width=0.8\linewidth, keepaspectratio]{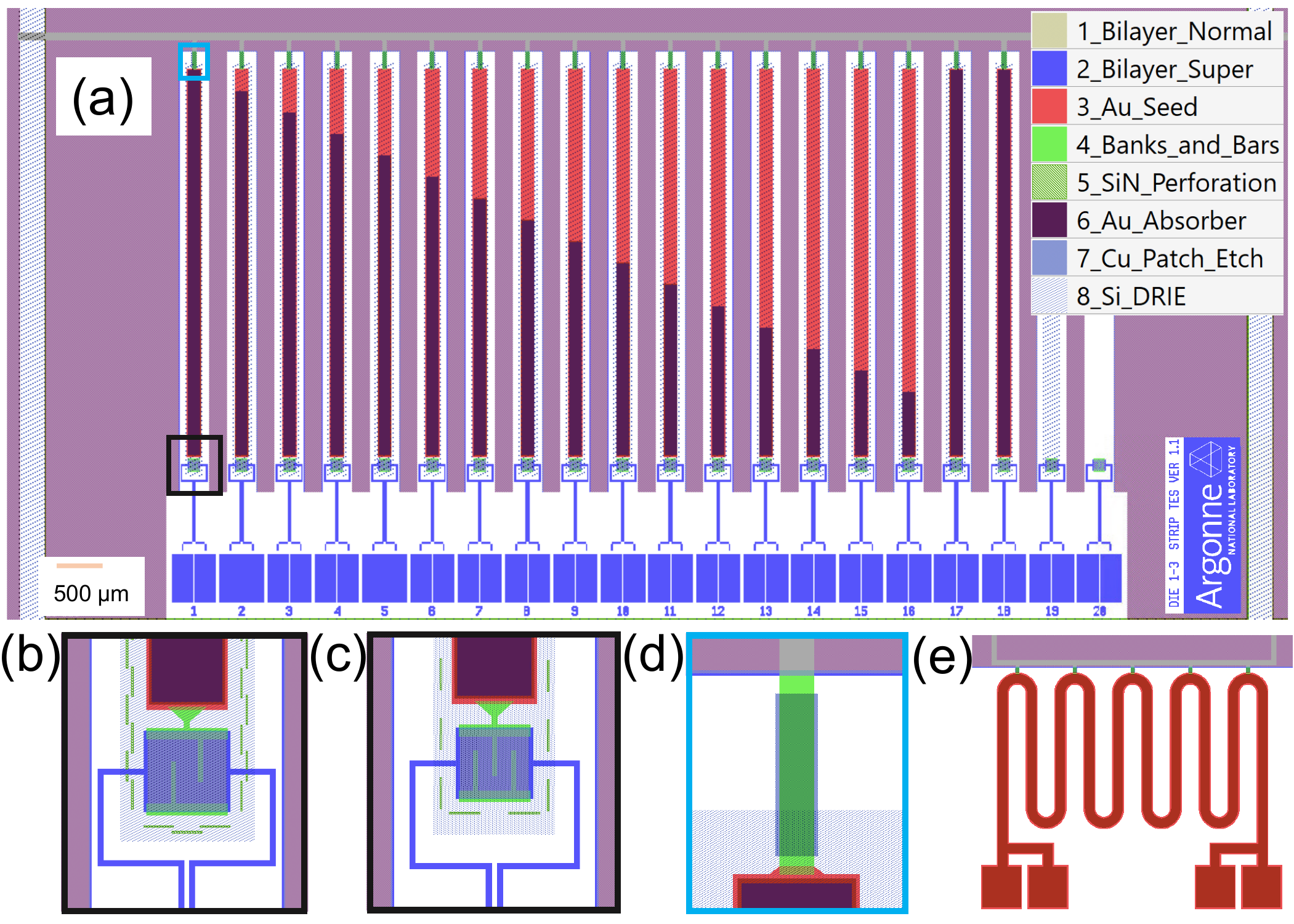}
\caption{\textbf{a} TES strip detectors with $20\times1$ linear array layout, die ($14\times7$) mm in size. The particular die is designed to vary the heat-capacity C while keeping G and $\alpha$ constant. Other variants of the design are in \textbf{b}, and \textbf{c}. \textbf{d} The design of the Cu patch for electroplating current path. \textbf{e} The design of the four probe test structure for measuring RRR of electroplated metal (Color figure online).}
\label{fig1}
\end{center}
\end{figure}

\begin{table}[htbp]
\caption{Typical layers for TES strip array fabrication}
\centering
\begin{tabular}{|l|l|l|l|}
\hline
Layer number & Material  & Process description               & Target thickness (nm) \\ \hline
0            & Mo/Cu     & Bilayer deposition                & 60/200                \\
1            & Cu        & Wet etch to define TES            & 200                   \\
2            & Mo        & Wet etch to define TES and wiring & 60                    \\
3            & Au        & Seed layer liftoff                & 100                   \\
4            & Cu        & Banks and bars liftoff            & 500                   \\
5            & SiNx/SiOx & Membrane perforation etch         & 500/160               \\
6            & Au        & Electroplating                    & 10k                   \\
7            & Cu        & Cu electroplating connection etch & 500                   \\
8            & Si        & Deep reactive ion etching         & 350k                  \\ \hline
\end{tabular}
\label{layers}
\end{table}

Standard low energy TESs are typically designed with a square geometry. For TESs, the geometry of the absorber can affect the X-ray pulse response and detection efficiency. Thus, the volume of the absorber is an important parameter which affects shape of the X-ray pulses because the pulse height $\propto$ 1/C, pulse decay time $\tau_0$ $\propto$ C, and energy resolution $\Delta\text{E}$ $\propto$~$\sqrt\text{C}$ \cite{Ullom15}. The strip detector was designed for a 3 inch wafer consisting of 24 dies with 20 devices on each die. Each die explores different aspects of T$_{\rm{c}}$, $\alpha$, G, and C. The TES itself is designed with a square geometry from a Mo/Cu bilayer with normal metal Cu banks and bars. The geometry of the SiN$_{\rm{x}}$ membrane and absorber is designed in a linear strip array fashion similar to the Ge strip detector typically used for EDXRD. TESs were designed to have up to a factor of two higher I$_{\rm{c}}$(0) than that of low energy (lower C) pixels in order to compensate for the effect of relatively high G of strip membrane geometry on $\alpha$ \cite{Wehle16, Morgan17}. Each device in a chip was provided with an electroplating connection at the top of the array, as shown in Fig.~\ref{fig1}a. In order to vary C, the device array was designed to vary plating area while keeping the seed layer over the full membrane for uniform heat flow out to the cold bath. The particular die was designed to study the effect of absorber geometry on X-ray pulse response. It also allows us to probe the effect of absorber area on heat flow to the cold bath. The plating area density was designed to keep relatively uniform throughout the wafer. For each die, devices 17-20 were designed to explore variation in T$_{\rm{c}}$ due to stress on membranes. These devices allowed for straightforward comparison of T$_{\rm{c}}$ on membrane with absorber and without absorber, and on solid substrate. We conducted our overall fabrication effort with a mask-less aligner which eliminated physical mask sets typically required from vendors and allowed us good control over the whole fabrication process. The typical fabrication process flow for the strip array is described in Table~\ref{layers}. In this report, we describe results from a die shown in Fig.~\ref{fig1}a. The other variants of the devices are currently under study.

\subsection{Mo/Cu Bilayer for TES}

\begin{figure}[htbp]
\begin{center}
\includegraphics[width=0.8\linewidth, keepaspectratio]{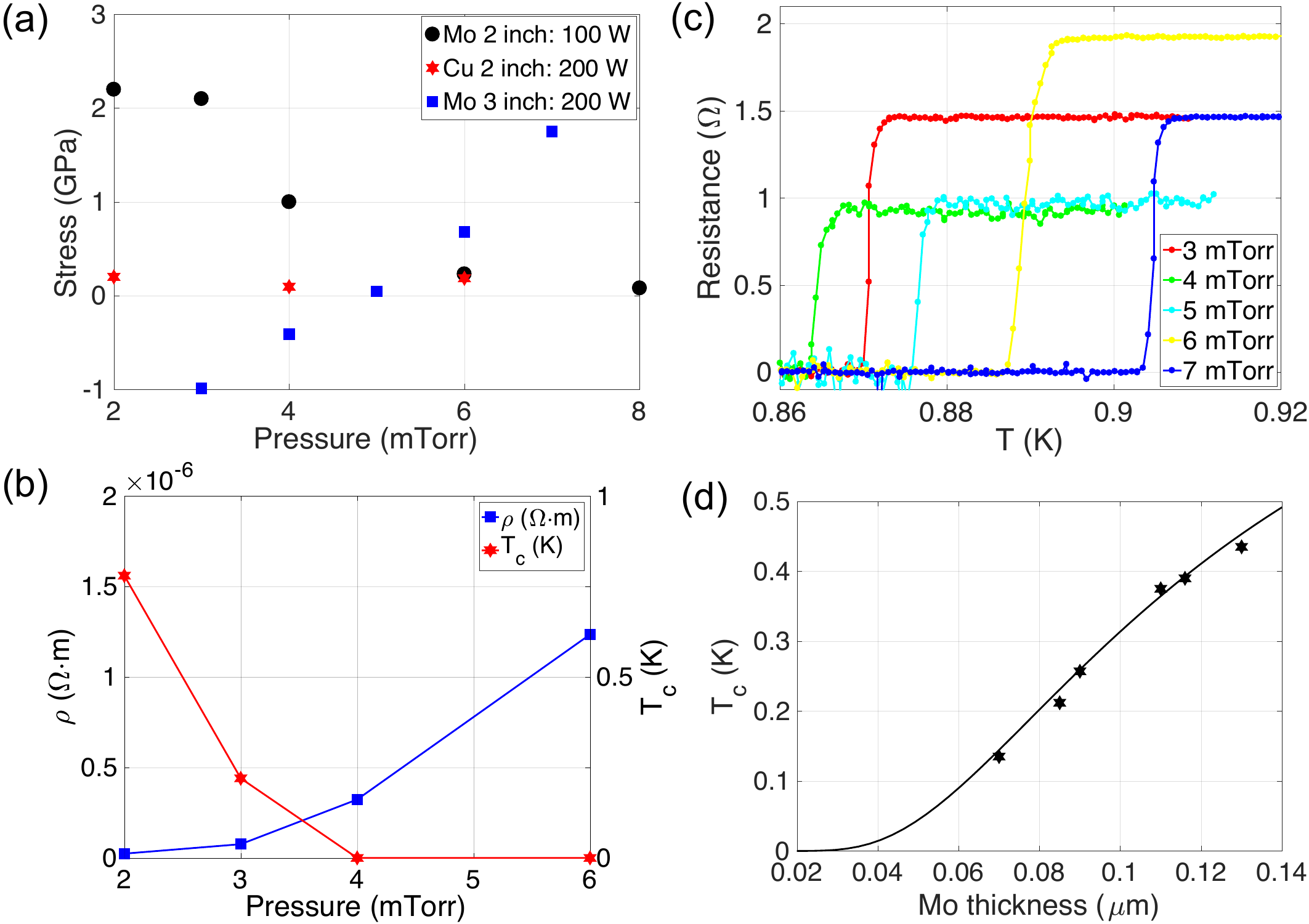}
\caption{\textbf{a} Residual stress in sputtered Mo and Cu thin films at different sputter powers, and gun geometry. \textbf{b} Resistivity $\rho$ and T$_{\rm{c}}$ of Mo films sputtered with 2 inch gun versus Ar gas pressure. \textbf{c} R versus T curves of the Mo films deposited with 3 inch sputter gun at 200~W power with stress described in Fig.~\ref{fig2}a. \textbf{d} Bilayer T$_{\rm{c}}$ as a function of different Mo thicknesses for a given Cu film thickness of $\sim\,200\,~\text{nm}$. $Black line$ is a fit function from Usadel theory for predicting the T$_{\rm{c}}$ of a thin normal metal-superconductor bilayer \cite{Martinis00}. The fit gives the interface transparency. (Color figure online)}
\label{fig2}
\end{center}
\end{figure} 

Robust, repeatable bilayer deposition process with a good control over T$_{\rm{c}}$ uniformity is needed for TES detectors. This required both higher energetics and confocal sputtering, as demonstrated later. We explored the deposition parameter space needed to understand residual stress in our films, and to optimize RRR, R$_{\rm{sq}}$, and T$_{\rm{c}}$ suitable for TES detectors. We describe here the steps taken to develop this process in our group. The Mo and Cu thin films were deposited using a DC magnetron sputtering tool consisting of four 2 inch sputter guns in a confocal geometry and one 3 inch central sputter gun. Intrinsic stress of the films was measured with a stylus profilometer. First, we studied the intrinsic stress of the films deposited using a 2 inch sputter gun at the closest possible working distance of 155~mm, as shown in Fig.~\ref{fig2}a. Clearly with 2 inch sputter gun, the intrinsic stress of Mo is very high for most of the Ar gas pressure range, whereas intrinsic stress of Cu is relatively low. For these Mo films, we see mostly tensile stress for whole Ar gas pressure range and deposition powers below 250 W.   
The cryogenic measurements of these films show either no signature of, or weak superconductivity, in the films down to 50~mK as shown in Fig.~\ref{fig2}b. We observed direct relationship of T$_{\rm{c}}$ with the resistivity $\rho$ of Mo films as shown in Fig.~\ref{fig2}b. This may point towards a weakly connected network of superconducting grains. We believe that the Mo films deposited by 2 inch sputter gun were not dense enough, and voids may exist in such films \cite{Windischmann92}. Thus, we further investigated films deposited with higher energetics such as higher power and with the central 3 inch sputter gun geometry. Films deposited with a relatively higher energy show a clear transition from compressive to tensile stress as shown in Fig.~\ref{fig2}a. We deposited our Mo films using parameters which give slightly compressive stress~\cite{Hilton01}. The cryogenic measurements of these films showed T$_{\rm{c}}$ close to bulk value ($\sim\,1$~K) as shown in Fig.~\ref{fig2}c. The RRR of Mo was $\sim\,2.2$ and Cu was $\sim\,6$ with a bilayer resistance of $\sim\,9~\text{m}\mathrm{\Omega}/\text{sq}$. We prepared our bilayer film by first depositing the Mo film at 4.5~mTorr, 200~W with a 3 inch sputter gun and then~\textit{in situ} Cu deposition at 4.5~mTorr, 200~W. The measured transparency of these bilayer films is $\sim\,0.22$ as shown in Fig.~\ref{fig2}d. In future work, we will explore the confocal 2 inch Mo gun at higher powers to improve film uniformity.

\subsection{Au Electroplating}
Electroplating for the thick Au absorber was done in a 2 L beaker. Two plating baths were used: 1.4 L Techni-Gold 25E Ready to Use (RTU) solution with a Au content of $\sim\,$~12.324 g/L, and Neutro\-nex-309 RTU solution with a Au content of $\sim\,$~10.28 g/L. Both solutions were heated to 40 $^{\circ}$C and 45 $^{\circ}$C respectively on a hot plate with a stir rate of 150 rpm to ensure uniform temperature throughout the solution. The plating current was supplied via a Keithley-2400 current source with a typical current density of 1~$\text{mA}/\text{cm}^{2}$ resulting in a plating rate of $\sim\,0.06~\mu\text{m}/\text{min}$ for both setups. SPR 220-7 photoresist (PR) was spin coated at 1500 rpm on a 3 inch wafer (PR thickness $\sim\,$$12~\mu\text{m}$) and used as a mold for $10~\mu\text{m}$ thick Au plating. The total area being plated per wafer was 2.534~$\text{cm}^2$ with a plating area density of 6.44$\%$. The films plated with Techni-Gold 25E generally have higher stress but smaller grains compared to the films plated with Neutronex-309. Thus, we plated first $1~\mu\text{m}$ of Au with Techni-Gold 25E followed by immediate DI water rinse and the rest of the thickness with Neutronex-309 to ensure no voids in the film, while keeping lower stress which would otherwise deform the mold as the film gets thicker. The average thickness variation of absorbers on a die was less than the surface roughness of the film $\sim\,$~400 nm. The electrical connection for the absorber electroplating is a 500~nm Cu patch prepared during liftoff of the banks and bars layer. These Cu patches provide direct electrical connection from the edge of a wafer to each device for all dies on the wafer simultaneously. Since the desired electroplated absorbers for the strip array will be rather thick, it could pose a challenge to remove these patches after absorber plating. For the current fab run, we accomplished this task by spin coating of $10~\mu\text{m}$ SPR 220-7 PR. We are currently investigating different approaches for patch removal for thicker absorbers.

\subsection{DRIE Process for Milli-meter Long SiN$_{\rm{x}}$ Strip Membranes}

\begin{figure}[htbp]
\begin{center}
\includegraphics[width=1\linewidth, keepaspectratio]{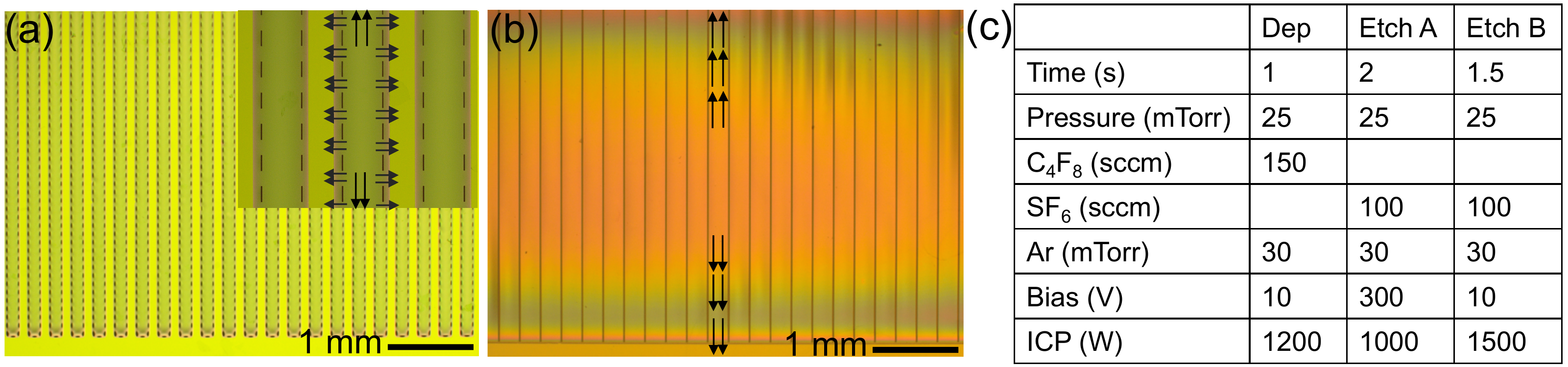}
\caption{Tests of long strip membranes for two different geometries: \textbf{a} Geometry I allows dominant controlled heat flow (black arrows) to the bath perpendicular to the strip, \textbf{b} Geometry II allows dominant controlled heat flow (black arrows) to the bath along the strip. \textbf{c} DRIE recipe for releasing the membranes (Color figure online).}
\label{fig3}
\end{center}
\end{figure}

We developed a robust DRIE etch process for releasing thin SiN$_{\rm{x}}$ membranes in our group. We utilized a Plasmatherm DRIE system to accomplish this task. We addressed several key issues such as wafer heating, micro-masking, etch selectivity, and device carrier suitability. A recipe optimized for low heat generation and suitable for making TES array is listed in Fig.~\ref{fig3}c. The ratio of selectivity for Si/PR is $\sim\,$~50. The Si etch rate was $\sim\,4~\mu\text{m}/\text{min}$ optimized for slower etch, and lower heat generation (below wax melting temperature $\sim\,$~105~$^{\circ}$C) for this recipe. We found that the standard sapphire carrier used in TES fabrication results in complete etch stop of Si due to polymer buildup in the bottom of the trench for all recipes tried including the one listed in Fig.~\ref{fig3}c. This could be due to a very different impedance of sapphire compared to Si causing much lower bias seen by the substrate (lower bombardment energy of the ions) for this system. We tried to compensate for this effect by reducing the polymer deposition time but it was not sufficient. Thus, we used a 4 inch Si wafer as a carrier wafer for our DRIE step. We mounted our wafer stack of 3 inch TES device wafer wax bonded to 3 inch sapphire on a 4 inch Si carrier. We observed minimal lateral Si etch ($\sim\,$~$10~\mu\text{m}$) for areas relevant to geometry I shown in Fig.~\ref{fig3}a with a sidewall slope of $\sim\,$~91.5$^{\circ}$. We used this recipe to conduct feasibility studies of the robustness of these membranes by themselves and as we added a large volume of electroplated absorber metals. Large membrane perimeter results in a large G. Thus, we tested two designs to control G; strips with width of $150~\mu\text{m}$ with membrane support from all four sides (geometry I) and strips with a width of $250~\mu\text{m}$ with membrane support from only two sides (geometry II). Geometry I resulted in a robust strip array while geometry II resulted in very good yield after membrane release but suffered during chip handling. Mechanical stability of both geometries can further be improved by preparing TESs on $1-2~\mu\text{m}$ thick SiN$_{\rm{x}}$ membranes, but with a cost of higher G. Geometry I was then electroplated with $10~\mu\text{m}$ Au and continued to show yield of 100\% on most dies. In future work, we will add thick Bi films on these membranes which will increase X-ray absorption efficiency while adding a moderate heat capacity.

\section{Results and Discussions}

\begin{figure}[htbp]
\begin{center}
\includegraphics[width=0.8\linewidth, keepaspectratio]{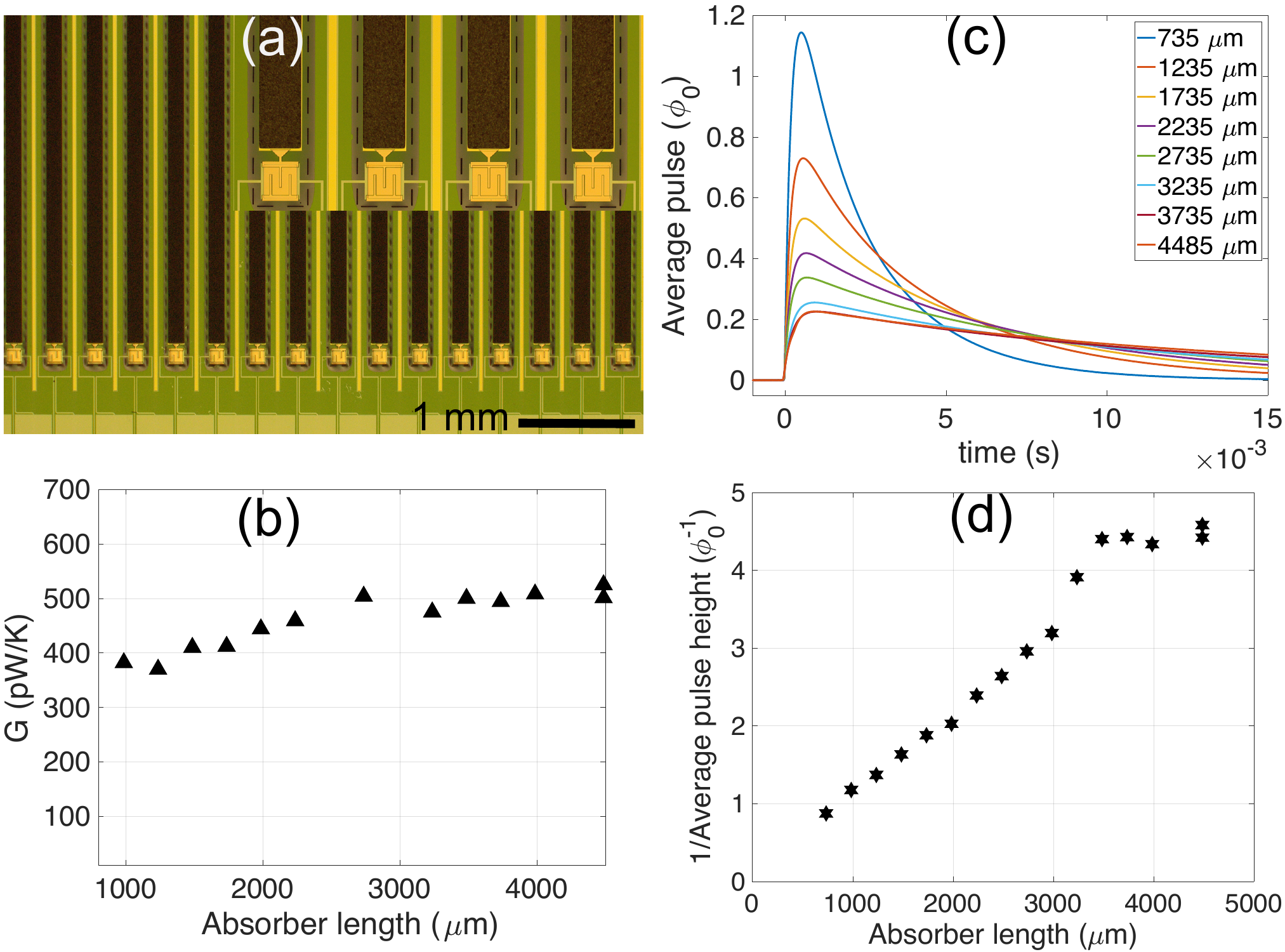}
\caption{\textbf{a} Micrograph of the fabricated strip TES linear array device. \textbf{b} Thermal conductance G \textbf{c} Average X-ray pulses and \textbf{d} Average pulse height determined from pulses in Fig.~\ref{fig4}c as a function of different absorber length for a die shown in Fig.~\ref{fig1}a . The value of unit $\phi_0$  $\sim\,8.96~\mu\text{A}$ (Color figure online.)}
\label{fig4}
\end{center}
\end{figure} 

We show a micrograph of the fabricated strip TES array in Fig.~\ref{fig4}a. The devices were cooled in an adiabatic demagnetization refrigerator with a base temperature of $\sim\,$~50 mK. We measured the I-V characteristics of the strip TES at different bath temperatures, and determined the Joule power dissipated in TES near T$_{\rm{c}}$ ($\sim\,$~80 mK) as a function of different bath temperatures using microwave SQUID multiplexers from NIST. We fit the experimental data to obtain thermal conductance G of all strip TES devices on a die shown in Fig.~\ref{fig1}a. In Fig.~\ref{fig4}b, the plot of G as a function of absorber length shows that G values are $\sim\,$~500 pW/K for all devices in an array suggesting the thermal conductance to the bath is nearly independent of the absorber area. Thus, the heat flow from the absorber to the cold bath is dominated by membrane geometry rather than by the absorber area. The observed variation in G ($<$~25$\%$) is still within the capability for the microwave-multiplexed readout and within required moderate count rate for the proposed application. Furthermore, the optimized device layout will be of a uniform strip length. For all strip TESs in an array, the average measured value of G is nearly a factor of 2.5 smaller than the designed value based on measurements done in standard square membrane devices \cite{Yan17}. We are currently investigating the origin of this, but this discrepancy could be indicative of a different phonon flow mechanism between two different membrane geometries. A comparable level of reduction in G was recently reported in additional layer of patterned normal metal Au features on SiN$_{\rm{x}}$ membranes which interfere with the phonon transport due to scattering by free electrons \cite{Zhang19}. Additionally, the strip TES pixels were irradiated with X-ray fluorescence generated from Mo foil with W-anode X-ray tube without collimator on the detector chip. The absorption efficiency for these devices with $\sim\,$$10~\mu\text{m}$ of Au is $\sim\,$80\% at 17 keV and drops to $\sim\,$40\% at 30 keV. We observed that strip TES pixels were minimally affected by the strip geometry. The pulses measured from a die shown in Fig.~\ref{fig1}a are plotted in Fig.~\ref{fig4}c. The inverse of pulse height (Fig.~\ref{fig4}d) and pulse decay time show a linear relationship with absorber length and thus scales with C, suggesting that the X-ray response of TESs is minimally affected by rather long strip geometry of the TES. For smaller pulses (absorber length $>$~$3500~\mu\text{m}$) in Fig.~\ref{fig4}d, relative uncertainty for pulse height determination was high. Initial estimates of the energy resolution, determined from fitting the Mo K$_{\rm{\alpha}}$ doublet, give a FWHM of approximately 180 eV and 40 eV for the longest ($4485~\mu\text{m}$) and shortest ($735~\mu\text{m}$) absorbers respectively. These estimates are close ($\sim\,$10\%) to the optimal filter predicted energy resolution based on the average pulse shapes and noise power spectral density. The difference in energy resolution could be due to the $\sim\,$6 times increased heat capacity and the position dependence in the longest absorber. Further measurements are needed to quantify their effects on energy resolution. Next generation devices will accommodate optimized heat capacity from electroplated Au/Bi stack to achieve the best energy resolution possible with reasonable absorption efficiency ($>$~20$\%$) at $\sim\,$100~keV.

\section{Conclusions}
We successfully designed, fabricated, and tested a strip TES linear array detector. We demonstrated that strip TESs designed for higher energies are minimally affected by rather long strip membrane, and absorber geometry. The X-ray pulses from strip TES showed a linear behavior for pulse height with different absorber lengths. In addition, the thermal conductance G to the bath was nearly independent of absorber area suggesting that heat flow to the bath depends on membrane geometry. Initial estimates showed energy resolutions of 40 eV to 180 eV near 17 keV for the absorber lengths studied. The measurement results from prototype 20-pixel strip TES detector show good promise for a path towards building an instrument tailored for energy dispersive X-ray diffraction imaging and Compton scattering applications. We plan to fabricate next generation devices with thick electroplated Au/Bi stack and study energy resolution, quantify effect of position dependence and obtain a proof-of-principle X-ray spectrum near 100 keV. 

\begin{acknowledgements}
This work was supported by the Accelerator and Detector R $\&$ D program in Basic Energy Sciences' Scientific User Facilities Division at the Department of Energy and the Laboratory Directed Research and Development program at Argonne. This research used resources of the Advanced Photon Source and Center for Nanoscale Materials (CNM), U.S. Department of Energy Office of Science User Facilities operated for the DOE Office of Science by the Argonne National Laboratory under Contract No. DE AC02 06CH11357. The authors gratefully acknowledge assistance from CNM staff, especially Dr. D. Czaplewski, and C.S. Miller. This work made use of the Pritzker Nanofabrication Facility (PNF) of the Institute for Molecular Engineering at the University of Chicago, which receives support from Soft and Hybrid Nanotechnology Experimental (SHyNE) Resource (NSF ECCS-1542205), a node of the National Science Foundation's National Nanotechnology Coordinated Infrastructure. The authors gratefully acknowledge assistance from PNF staff, especially P. Duda and A. Mukhortova. We thank Dr. O. Makarova from Creatv MicroTech, Inc. for discussions on Au electroplating. We also thank Dr. D. Schmidt from the Quantum Sensors Group at NIST for discussions on TES fabrication.

\end{acknowledgements}

\end{document}